\documentclass[journal,comsoc]{IEEEtran}

\usepackage[dvips]{graphicx}
\usepackage{latexsym}
\usepackage{epsfig}
\usepackage{color}
\usepackage{amsmath} 
\usepackage{amssymb}
\usepackage{amsxtra}
\usepackage{enumitem}

\usepackage{units}

\usepackage[T1]{fontenc}
\usepackage{setspace}

\usepackage{amsmath}
\usepackage[cmintegrals]{newtxmath}
\usepackage{graphicx}
\graphicspath{{Figures/}}
\usepackage{url}
\usepackage{cite}
\usepackage{algorithm}
\usepackage{algpseudocode}

\makeatletter
\def\BState{\State\hskip-\ALG@thistlm}
\makeatother

\usepackage{epstopdf}
\usepackage{amsmath}
\usepackage{amsxtra}
\usepackage{amstext}
\usepackage{amssymb}
\usepackage{latexsym}
\usepackage{dsfont} 
\usepackage{color}
\usepackage{multirow}
\usepackage{float}
\usepackage{hyperref}

\usepackage{tabularx,colortbl}
\usepackage[table]{xcolor}

\usepackage{lscape}

\usepackage{algorithm}
\usepackage{algpseudocode}

\usepackage{units}
\usepackage{cases}
\usepackage[ subrefformat=parens,labelformat=parens, caption=false, font=footnotesize]{subfig}
\usepackage{mathtools}
\mathtoolsset{showonlyrefs} 

\usepackage{mathtools}

\newcommand{\maxx}{\mathrm{max}}
\newcommand{\T}{\mathrm{T}}

\begin{document}

\title{Next Generation Terahertz Communications: A Rendezvous of Sensing, Imaging, and Localization}

\author{Hadi~Sarieddeen,~\IEEEmembership{Member,~IEEE,}
        Nasir~Saeed,~\IEEEmembership{Senior Member,~IEEE,}
        Tareq~Y.~Al-Naffouri,~\IEEEmembership{Senior Member,~IEEE}
        and~Mohamed-Slim~Alouini,~\IEEEmembership{Fellow,~IEEE}
\thanks{The authors are with the Division of Computer, Electrical and Mathematical Sciences and Engineering, King Abdullah University of Science and Technology (KAUST), Thuwal 23955-6900, Saudi Arabia (e-mail: \{hadi.sarieddeen; nasir.saeed; tareq.alnaffouri; slim.alouini\}@kaust.edu.sa).
}
\thanks{This work was supported by the King Abdullah University of Science and Technology (KAUST) Office of Sponsored Research.}
}

\maketitle

\begin{abstract}

Terahertz (THz)-band communications are celebrated as a key enabling technology for next-generation wireless systems that promises to integrate a wide range of data-demanding and delay-sensitive applications. Following recent advancements in optical, electronic, and plasmonic transceiver design, integrated, adaptive, and efficient THz systems are no longer far-fetched. In this paper, we present a progressive vision of how the traditional ``THz gap'' will transform into a ``THz rush'' over the next few years. We posit that the breakthrough that the THz band will introduce will not be solely driven by achievable high data rates, but more profoundly by the interaction between THz sensing, imaging, and localization applications. We first detail the peculiarities of each of these applications at the THz band. Then, we illustrate how their coalescence results in enhanced environment-aware system performance in beyond-5G use cases. We further discuss the implementation aspects of this merging of applications in the context of shared and dedicated resource allocation, highlighting the role of machine learning.

\end{abstract}

\IEEEpeerreviewmaketitle

\section{Introduction}
\label{sec:Intro}

Wireless communication carrier frequencies have been gradually expanding over recent years in an attempt to satisfy ever-increasing bandwidth demands. Since the terahertz (THz) band is the last unexplored band of the radio frequency (RF) spectrum, technologies from both the neighboring microwave and optical bands are being explored to support THz communications. Due to the lack of compact and efficient THz devices (the so-called ``THz gap''), THz-band applications have been traditionally restricted to the areas of imaging and sensing. However, following recent advancements in THz signal generation, modulation, and radiation, communication-based THz-band use cases can now be foreseen \cite{elayan2019terahertz}. 

Recent advancements in THz transceiver designs are mainly electronic and photonic. While photonic technologies have a data rate advantage, electronic platforms are superior in their ability to generate higher power \cite{sengupta2018terahertz} ($\unit[100]{\mu W}$ to $\unit[]{mW}$ compared to typical tens of $\unit[]{\mu W}$ in photonics). Electronic solutions \cite{Kenneth8808165} based on silicon complementary metal-oxide-semiconductor (CMOS) technology can generate up to $\unit[1.3]{THz}$ signals. The corresponding highest unity current gain frequency ($f_{\T}$) and unity maximum available power gain frequency ($f_{\maxx}$), however, remain at 280 gigahertz (GHz) and $\unit[320]{GHz}$, respectively. On the photonic side \cite{sengupta2018terahertz}, beyond-$\unit[300]{GHz}$ frequencies have been supported using uni-traveling carrier photodiodes, photoconductive antennas, optical down-conversion systems, and quantum cascade lasers. Furthermore, since satisfying emerging system-level properties requires designing efficient and programmable devices, as opposed to designing best-performing THz devices, integrated hybrid electronic-photonic systems are emerging \cite{sengupta2018terahertz}. In particular, systems composed of photonic transmitters and III-V compound semiconductor-based electronic receivers have demonstrated very high data rates. Furthermore, plasmonic solutions based on novel materials such as graphene \cite{Jornet5995306} are also gaining popularity. Plasmonic antennas support surface plasmon polariton waves over the metal-dielectric interface. They possess high electron mobility and tunability, which results in reconfigurable and compact array designs. Following these advances, it is clear that the THz gap is rapidly closing.



%

THz communications are attracting both praise and criticism. Is pushing microwave communications beyond the well-established millimeter-wave (mmWave) band worth the effort? And why should we settle for THz communications if high data rates can be supported by the more mature visible light communications (VLC)? In this paper, we discuss how THz communications can reap the benefits of both mmWave and VLC. We argue that the peculiarities of THz sensing and imaging, combined with the high-resolution localization capabilities, can significantly boost the performance of future communication systems and introduce a plethora of novel applications.  We advocate this merging of applications by addressing its implementation aspects and demonstrating proof-of-concept simulations; we show that THz sensing and localization can be seamlessly piggybacked onto THz communications. We start by treating each of these applications independently. 


\begin{figure*}[t]
  \centering
  \includegraphics[width=0.88\textwidth]{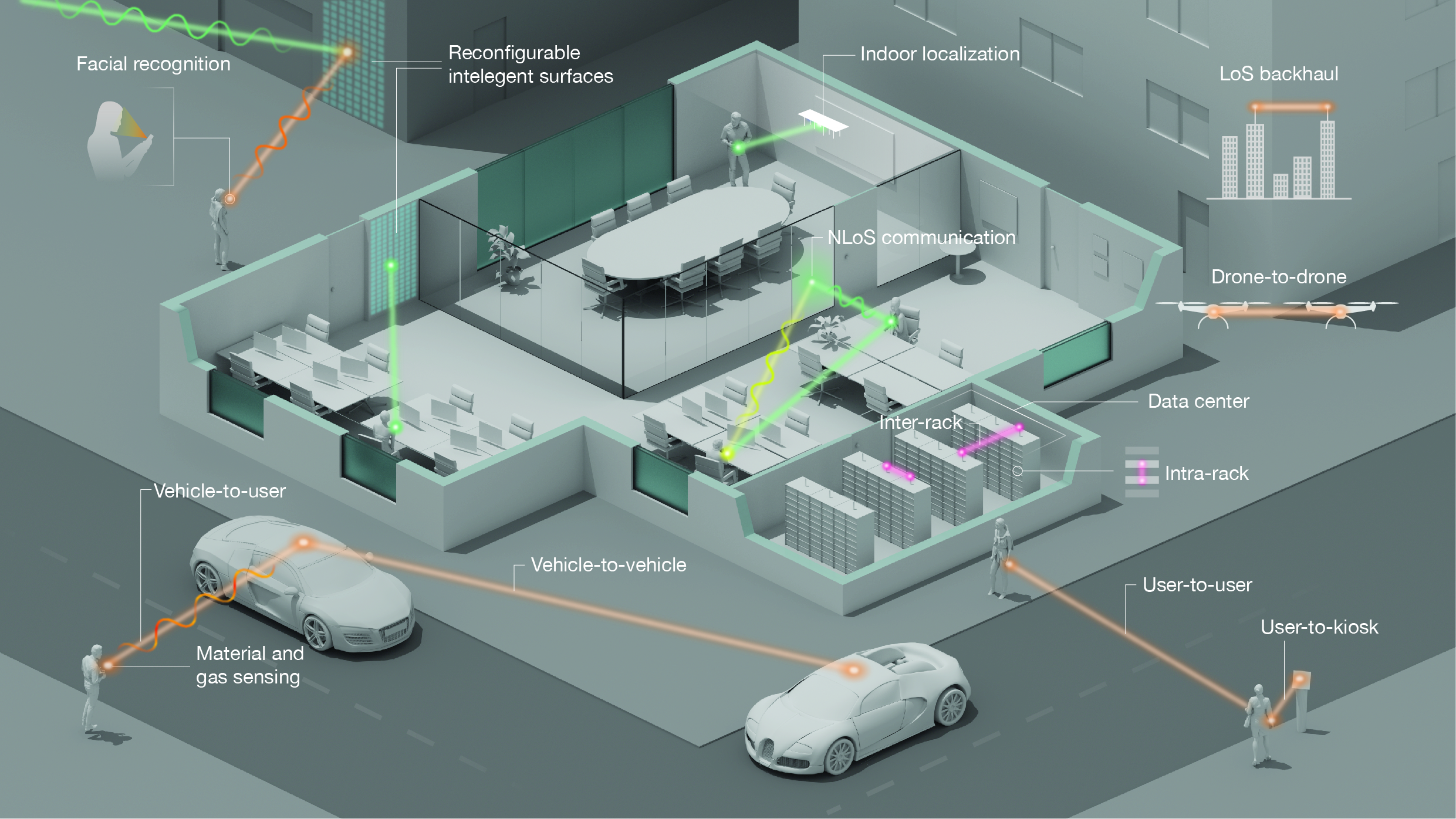}
 \caption{Prospective indoor and outdoor applications of THz communications (the figure is created by Ivan Gromicho, Scientific Illustrator at KAUST).}
  \label{Fig_apps}
\end{figure*}

\section{THz Communications}
\label{sec:Commun}

THz-band communications are expected to play a pivotal role in the upcoming sixth-generation (6G) of wireless mobile communications, enabling ultra-high bandwidth communication paradigms. To this end, many research groups have attracted significant funding to conduct THz-related research and standardization efforts have been launched \cite{elayan2019terahertz}. Despite the fact that by definition the THz band extends from $\unit[300]{GHz}$ to $\unit[10]{THz}$, researchers have found it convenient to categorize beyond-$\unit[100]{GHz}$ applications as THz communications, below which the millimeter-wave bands of 5G are defined.

\subsection{An Argument for THz Communications}
\label{sec:argument}

Unlike mmWave communications, THz communications can leverage the large available bandwidths at the THz band to achieve a terabit/second data rate without additional spectral efficiency enhancement techniques \cite{Jornet5995306}. Pushing research beyond mmWaves is thus inevitable, as some applications can never be accomplished in any current or envisioned mmWave system (transmitting an uncompressed 8K-resolution video, for example). Due to their shorter wavelengths, THz communication systems can support higher link directionality, are less susceptible to free-space diffraction and inter-antenna interference, can be realized in much smaller footprints, and possess higher resilience to eavesdropping. Furthermore, the THz band promises to support higher user densities, higher reliability, less latency, more energy efficiency, higher positioning accuracy, better spectrum utilization, and increased adaptability to propagation scenarios. 

On the other side of the spectrum, optical communications over the infrared and visible bands, mainly VLC, are very mature (except for the ultraviolet band, which has limitations). VLC has the potential to provide higher data rates than the rates promised by THz communications, and at a low cost. However, the two technologies are fundamentally different \cite{elayan2019terahertz}. For instance, THz signals are not affected by ambient light, atmospheric turbulence, scintillation, cloud dust, or temporary spatial variations of light intensity in the same way that optical signals are affected. More importantly, the severe unguided narrow-beam properties of optical signals require more delicate pointing, acquisition, and tracking to guarantee alignment. Despite possessing quasi-optical propagation traits, THz communications retain several characteristics of microwave communications, and they can still draw on reflections and antenna array-processing techniques to support non-line-of-sight (NLoS) propagation and efficient beamforming, respectively. Nevertheless, due to human eye sensitivity, VLC is not the best option for uplink communications,  and it is susceptible to a different type of blockage. Therefore, availability can be enhanced by using both mmWave/THz communications and VLC in a heterogeneous setup.

\subsection{Challenges and Solutions}
\label{sec:challenges}

Many challenges need to be addressed prior to the widespread introduction of THz communications. For instance, the THz band's high propagation losses and power limitations result in very short communication distances, and frequency-dependent molecular absorptions result in band-splitting and bandwidth reduction. Signal misalignment and blockage are also more severe at the THz band. Furthermore, except for recent measurements at sub-THz frequencies \cite{Rappaport8732419}, there are no realistic THz channel models. In fact, the THz channel is dominated by a line-of-sight (LoS) and a few NLoS paths because of significant signal reflection losses and very high signal diffraction and scattering losses. Note that NLoS communication can naturally exist at the THz band because of the relatively low loss in specular reflection from specific surfaces (drywall for example). We hereby summarize some signal processing solutions to overcome these challenges (see \cite{Sarieddeen8765243,faisal2019ultra} and references therein).

\subsubsection{UM-MIMO and RIS systems}
\label{sec:UM-MIMO}

Very dense ultra-massive multiple-input multiple-output (UM-MIMO) antenna systems can provide the required beamforming gains necessary to overcome the distance problem. While mmWave UM-MIMO systems require footprints of a few square centimeters for a small number of antenna elements, a large number of antenna elements can be embedded at THz in a few square millimeters. Note that both the magnitude of the propagation loss and the number of antennas that can be fit in the same footprint increase in the square of the wavelength. Therefore, higher beamforming gains due to increased compactness can account for the increasing path loss. THz array-of-subarrays (AoSA) configurations enable hybrid beamforming, which in turn reduces hardware costs and power consumption and provides the flexibility to trade beamforming and multiplexing, for a better communication range and spectral efficiency, respectively. Configurable AoSA architectures can support multi-carrier communications and variations of spatial and index modulation schemes \cite{Sarieddeen8765243}, by tuning each antenna element to a specific frequency, turning it on/off, or changing its modulation type in real-time (in plasmonic solutions, the frequency of operation can be tuned by simple material doping or electrostatic bias). To enhance the correlated MIMO channel conditions in LoS scenarios, spatial tuning techniques \cite{Sarieddeen8765243} that optimize the separations between antenna elements can be applied. Nevertheless, a high multiplexing gain can only be achieved when the communication range is less than the so-called Rayleigh distance. Reconfigurable intelligent surfaces (RISs) can extend the communication range and the Rayleigh distance of the THz system.

\subsubsection{Waveform and modulation}
\label{sec:wav_mod}

Since bandwidth reduction from molecular absorption gets more pronounced at larger distances, the available absorption-free spectral windows are distance-dependent. Hence, distance-aware solutions that optimize waveform design, transmission power, and beamforming are required. Optimized THz-specific multi-carrier waveform designs other than orthogonal frequency-division multiplexing (OFDM) are favored. OFDM is very complex to implement at the THz band and it results in a peak-to-average power ratio problem; single-carrier schemes are recommended (alongside carrier aggregation) due to the large available bandwidths and flat nature of the channel. Moreover, since generating continuous modulations in compact architectures is challenging,  plasmonic pulse-based architectures (asymmetric on-off keying modulations that consist of transmitting femtosecond-long short pulses) are proposed \cite{Jornet5995306}. However, the output power of pulse-based systems is limited. In addition, since the frequency response of pulse-based systems covers the entire THz range, it is no longer possible to operate in absorption-free spectra. Re-transmissions following molecular absorption (re-radiation of absorbed energy with negligible frequency shifts) result in a frequency-dependent (colored) noise component that is induced by the channel.

\subsubsection{Low-complexity baseband}
\label{sec:baseband}

Many more challenges need to be addressed from a signal processing perspective to overcome the mismatch between the bandwidth of the THz channel (terabit/second) and that of the digital baseband system (which is limited by clock speeds of few GHz). Channel coding is the most computationally-demanding component of the baseband chain; nevertheless, the complete chain should be made efficient and parallelizable. Therefore, joint architecture and algorithm co-optimization of channel coding, channel estimation, and data detection is required. Low-resolution digital-to-analog conversion systems can further reduce the baseband complexity; all-analog THz solutions are also considered.

\section{THz Sensing and Imaging}
\label{sec:Sensing}

THz signals have traditionally been used for wireless sensing and imaging in applications such as quality control, food safety, and security. Several THz-wave propagation properties, which microwave and infrared waves lack, make THz signals a good candidate for material and gas sensing. In particular, many biological and chemical materials exhibit unique THz spectral fingerprints. THz signals can penetrate a variety of non-conducting, amorphous, and dielectric materials, such as glass, plastic, and wood. Moreover, metals strongly reflect THz radiation, which enables detecting weapons. Due to strong molecular coupling with hydrogen-bonded networks, THz signals can be used to observe water dynamics. Similarly, due to unique spectral signatures that arise from transitions between rotational quantum levels in polar molecules, THz signals can be used for gas detection (rotational spectroscopy). 

The most popular THz sensing method is THz time-domain spectroscopy (THz-TDS) \cite{jepsen2011terahertz}. THz-TDS is a pulse-based passive-sensing technique that probes a sample with THz broadband radiation in the form of short pulses and records the temporal profiles of these pulses with and without sample presence; the ratio of the electric field strengths of these profiles determines the optical properties of the sample material. There are two types of THz-TDS systems: transmission-based and reflection-based. While transmission spectroscopy analyzes the amount of light absorbed by a material, reflection spectroscopy studies the reflected or scattered light. Most THz sensing and imaging applications are conducted in transmission mode, since absorbance spectroscopy has an easier set-up design and results in higher signal contrast. Nevertheless, several factors make the reflection mode more appealing. For instance, reflection geometry enables detecting targets on non-transparent substrates and measuring the spectrum of highly absorptive materials (THz-opaque). More importantly, from a communication systems perspective, hand-held devices by themselves can support imaging/sensing in reflection mode without requiring infrastructure assistance.

\begin{figure}[t]
  \centering
  \includegraphics[width=0.45\textwidth]{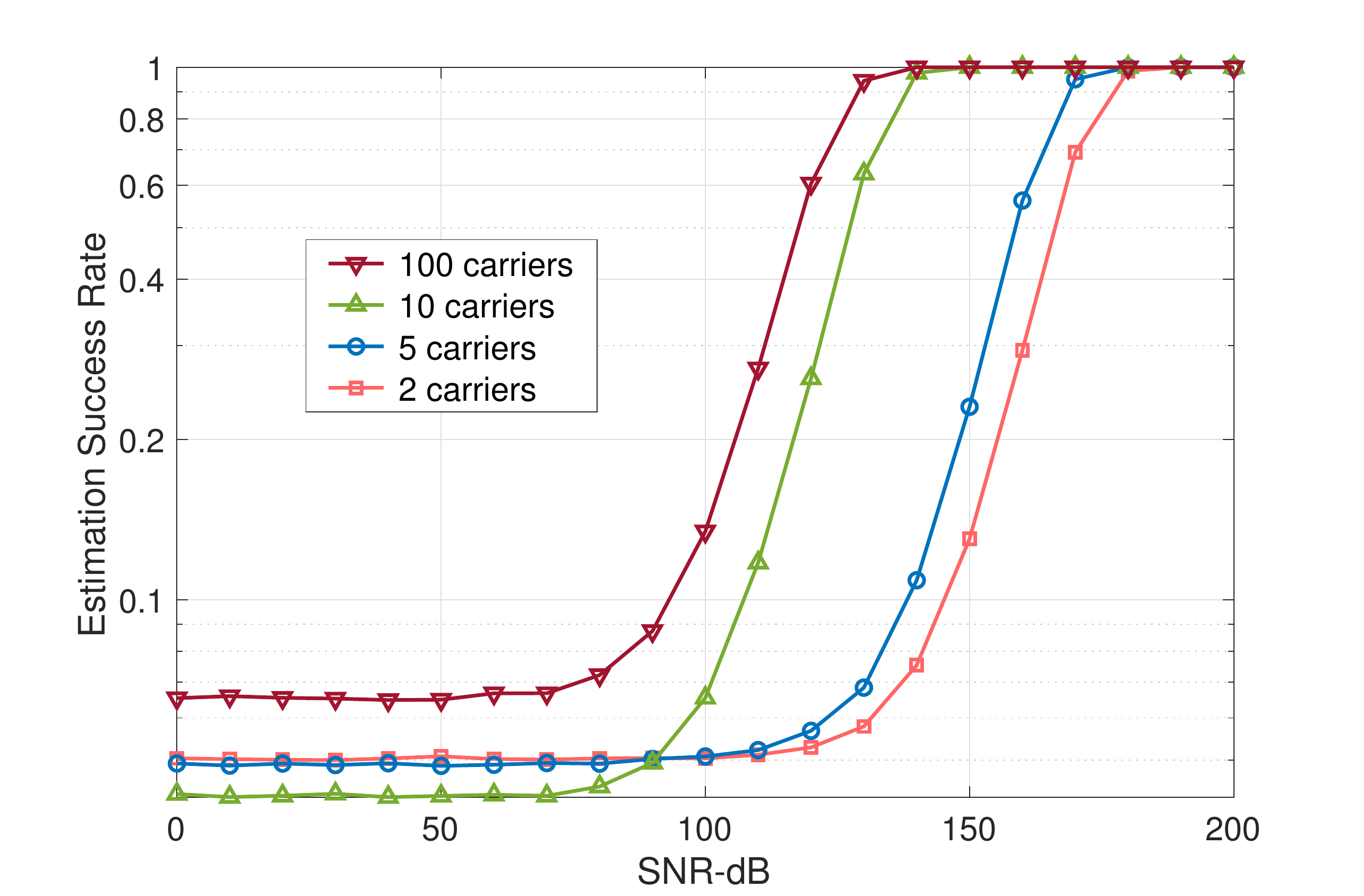}
 \caption{Performance analysis of carrier-based water vapor concentration sensing (accuracy $1\%$) over a $\unit[5]{m}$ range as a function of SNR and number of THz carriers (uniformly distributed between $\unit[1]{THz}$ and $\unit[2]{THz}$).}
  \label{Fig_gas}
\end{figure}



As an extension to sensing, THz-TDS imaging extracts spectral information to determine the type and shape of the material under study. THz-TDS imaging is amplitude-based, phase-based, or a combination of the two. It consists of measuring the waveform of a THz wave traversing or reflected from an object at multiple positions. Due to low scattering, THz-TDS imaging systems are capable of producing high-contrast images. In fact, mmWave and THz imaging techniques are more powerful than their infrared counterparts, since they are less sensitive to weather and ambient light conditions. The vastly wider channel bandwidths of THz can easily support imaging applications with large fields of view. Furthermore, the high gain of directional antennas enables focused directional sensing/imaging and much finer spatial resolution (sub-millimeter spatial differentiation \cite{Rappaport8732419}). 



With the advent of THz technology, carrier-based sensing/imaging systems can offer greater flexibility and fine-tuning capabilities over the frequency ranges of interest. In the following, we illustrate a proof-of-concept simulation for carrier-based THz-band wireless gas sensing (electronic smelling \cite{Kenneth8808165}). We assume a setup that consists of a transmitter shooting a few select high-frequency signals into a medium and a receiver estimating the channel response in order to detect absorption spikes. The estimated channel responses are correlated with a reference database on molecular absorption spectra (high-resolution transmission database (HITRAN) \cite{Jornet5995306}), and a decision is made on the gaseous constituents of the medium. For the specific case of measuring the percentage of water vapor, Fig. \ref{Fig_gas} illustrates the success rate versus the signal-to-noise ratio (SNR). A performance enhancement of $\unit[50]{dB}$ is noted by increasing the number of carriers from 2 to 100. However, an SNR of more than $\unit[100]{dB}$ is still required to counter the losses at $\unit[5]{m}$. This can be mitigated via antenna and beamforming gains ($\unit[30]{dB}$ beamforming gain with 1024-element subarrays). Note that, for this simulation, a heuristic search-based algorithm is assumed. To further enhance performance in more complex sensing scenarios, the collected THz measurements can be processed using filtering and noise-reduction schemes followed by rigorous feature-extraction and classification algorithms.

\section{THz Localization}
\label{sec:Localization}

\begin{figure*}
  \centering
  \includegraphics[width=0.97\textwidth]{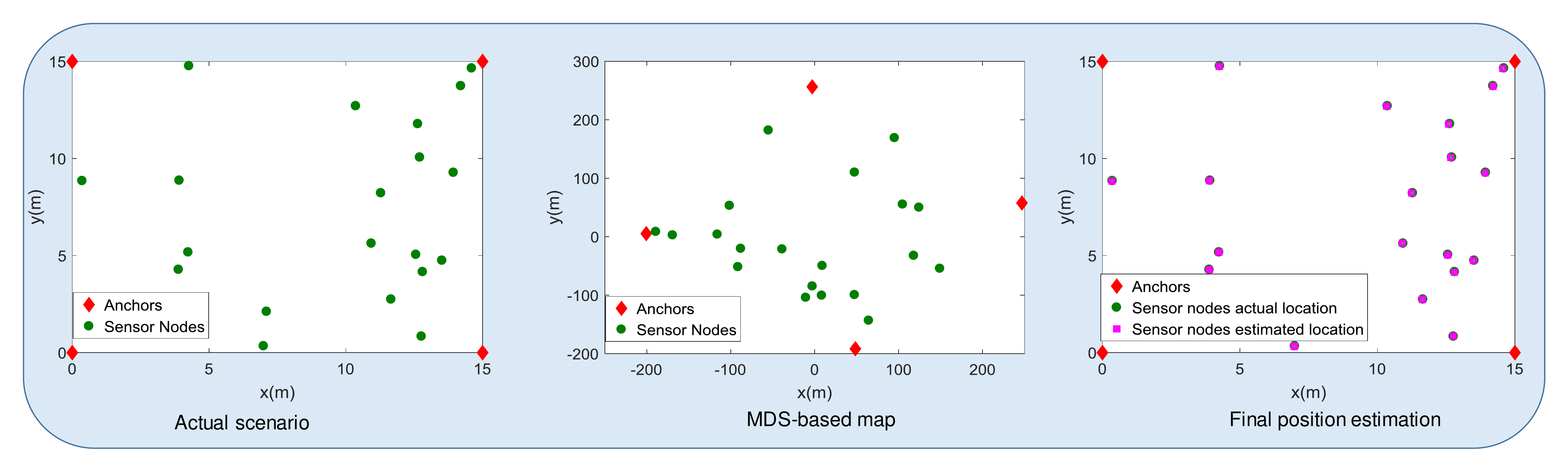}
 \caption{Localization in THz communication systems using MDS.}
  \label{Fig_locmds}
\end{figure*}

Next-generation wireless networks are expected to provide accurate location-based services. It is envisioned that 6G networks based on mmWave and THz frequencies will provide centimeter-level accuracy \cite{Rappaport8732419} that conventional GPS and cell multilateration-based localization techniques fail to provide. High-frequency localization techniques are based on the concept of simultaneous localization and mapping (SLAM), in which the accuracy improves by collecting high-resolution images of the environment; the THz band can provide such high-resolution images. SLAM-based techniques consist of three major steps: imaging of the surrounding environment, estimation of ranges to the user, and fusion of images with the estimated ranges. For instance, a sub-centimeter level of accuracy can be achieved by constructing three-dimensional (3D) images of the environment using signals between $\unit[200]{GHz}$ and $\unit[300]{GHz}$ and projecting the angle- and time-of-arrival information from the user to estimate locations. Since SLAM deals with relatively slow-moving objects, there is sufficient time to process high-resolution THz measurements. Such measurements can hold sensing information, resulting in complex state models comprising the fine-grained location, size, and orientation of target objects, as well as their electromagnetic properties and material types.


Besides SLAM-based techniques, little effort has been made to develop THz-specific localization methods. In \cite{Absi2018}, a weighted least-squares estimator is used for the localization of transceivers operating at the THz band. Tags with dielectric resonators are used as anchors to transmit beacon signals, whereas the receiver computes the round-trip time of flight (RToF) to the available anchors. The RToF ranging technique is used to avoid the synchronization issue in time-based ranging methods. By increasing the bandwidth of the pulse and the size of the receiver antenna, millimeter-level accuracy is demonstrated. Nevertheless, a direction-of-arrival (DoA)-based approach is studied in \cite{Shree2018} for ranging in nanoscale sensor networks operating at $\unit[6]{THz}$. A uniform linear array is used as a sink node with $N$ antennas, which estimates the direction of the THz signal arriving at its various antennas from different directions. The sensor node uses the multiple signal classification (MUSIC) algorithm to estimate the DoA measurements, which, at a distance of $\unit[6]{m}$, results in an accuracy of less than a degree.



Research on robust and accurate localization algorithms for future THz communications systems is still lacking (the aforementioned schemes focus on ranging only). We hereby consider THz network localization using multidimensional scaling (MDS), which is a data visualization technique. Let the input to the MDS algorithm be the pairwise distance between two arbitrary nodes, which is estimated using any of the above ranging schemes for THz communications. We assume that the estimated distance between two sensor nodes is corrupted by the distance-dependent additive Gaussian noise. Based on the pairwise estimated distances, the MDS algorithm tries to locate the sensor nodes in a given dimensional space. For example, consider a scenario in which $20$ sensor nodes are randomly distributed over a $\unit[15 \times 15]{m^2}$ area and $4$ anchor nodes are located at the corners of this area. MDS tries to graphically show the relationship between all the nodes in the network, as illustrated in Fig.~\ref{Fig_locmds}. The left plot of Fig.~\ref{Fig_locmds} shows the actual location of sensor nodes and anchor nodes, while the middle plot shows the MDS-based initial map of the network, which has no actual coordinates (it is centered around the origin). The right plot shows the final node estimation with an accuracy of $\unit[2.4]{cm}$ for a ranging-error variance of $\unit[8]{cm}$.


\section{6G and Beyond Use-Cases}
\label{sec:6G}

Having detailed the peculiarities of THz communications, sensing, and localization, we next envision prospective 6G-and-beyond THz-band use cases. Finland's 6G Flagship envisions THz technology to be a key driver for 6G ubiquitous wireless intelligence \cite{Matti_Oulu}. Through high-data-rate wireless remoting of human cognition, information shower, accurate localization, sensing, and imaging, a plethora of services across multiple industry verticals can thrive in many areas, including transportation systems, robotics, augmented reality, entertainment, and health care (see \cite{Rappaport8732419} and references therein). The large THz bandwidths and massive antenna arrays, combined with the inherent densification caused by machine-type communications, will not only result in enhanced communication-system performance, but also in enhanced sensing, imaging, and localization. For instance, array signal processing techniques such as 3D beamforming, which was originally proposed to extend communication ranges and enhance spectrum efficiency, can also enable precise positioning and tracking, as well as create images of physical spaces by systematically monitoring signals at a wide array of angles. 





While high-capacity THz links have already been advocated to replace wired backbone connectivity in network backhauls and data centers, the holy grail of THz communications is to enable indoor and outdoor mid-range mobile wireless communications. The following are some examples.




\subsection{Vehicular and Drone-to-Drone Communications}
\label{sec:vehicular}

Next-generation vehicular networks demand high data rates and reliable communications. For instance, see-through vision and bird's eye view require a data rate of $\unit[50]{Mbps}$ and a latency of $\unit[50]{ms}$. Similarly, automated overtake demands a $\unit[10]{ms}$ delay and $99.999\%$ reliability. Despite the uncertainty of the THz channel, high reliability can be achieved with high bandwidths and sufficiently-dense networks, as demonstrated in \cite{Chaccour8763780} for the case of THz virtual reality. THz-band communications are thus envisioned to be a key technology that can satisfy the reliability and real-time traffic demands of future vehicular networks \cite{Choi2016}, to enhance safety solutions and enable new applications, such as platooning and remote driving. Furthermore, THz sensing/imaging and localization/navigation are particularly useful in a multitude of vehicular communications applications. 



High-speed drone-to-drone communications can also be achieved at the THz band. Drones can form flying ad hoc networks (FANETs) to enable broadband communication over a large-scale area. In FANETS, the THz band achieves high capacity at greater flexibility compared to FSO, which has stringent pointing and acquisition requirements. Nevertheless, drones operating at the THz band would require millimeter levels of positioning accuracy \cite{Mendrzik2018}, as opposed to less than a centimeter for a drone operating at $\unit[30]{GHz}$.


\begin{figure*}[t]
  \centering
  \includegraphics[width=0.88\textwidth]{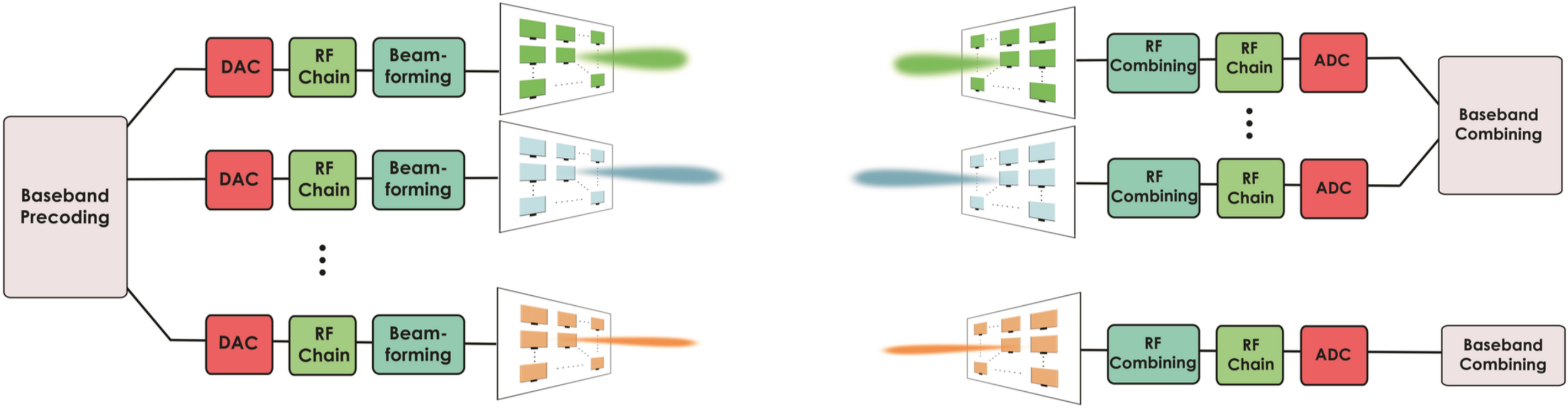}
 \caption{A typical THz communication system model featuring AoSA configurations and hybrid signal processing, with the digital and analog domains being separated by digital-to-analog conversion (DAC) and analog-to-digital conversion (ADC) blocks. }
  \label{Fig_RF}
\end{figure*}

\subsection{Reflection-Assisted Communications}
\label{sec:RIS}

In addition to the aforementioned applications, Fig. \ref{Fig_apps} illustrates the role of both regular (non-reconfigurable) specular reflecting surfaces and RISs. In general, the strong specular component makes existing building surfaces behave like electrical mirrors at sufficiently short THz wavelengths. Nevertheless, intelligent surfaces can be custom built from discrete-element semiconductors and metamaterials. Such RISs can scale up signal power and reflect THz signals towards a direction of interest by introducing specific phase shifts. An electrically-large RIS that supports all these functionalities can be achieved via relatively small footprints at high frequencies \cite{ntontin2019reconfigurable}. 


THz imaging, sensing, and localization can enhance the performance of RISs. For instance, the type of material that makes up two neighboring walls can be sensed to help determine the better NLoS route. Note that the change in signal color (Fig. \ref{Fig_apps}) after reflection indicates a change in signal characteristics (due to the reflection coefficient); this can be detected in order to classify the material type of the surface. Similarly, building real-time environment maps results in intelligent beamforming that makes use of reflecting surfaces (beamforming with environment awareness). Furthermore, the unreliability of the channel at high frequencies can be circumvented by sensing the environment in real time and finding alternative propagation routes. Conversely, RISs can enhance the performance of imaging, sensing, and localization, where seeing around corners in NLoS can be leveraged for rescue and surveillance applications as well as for autonomous localization and navigation.

\section{Implementation Aspects}
\label{sec:Implementation}

Given that research on THz communications is still in its infancy, the applications discussed in Sec. \ref{sec:6G} may seem futuristic. In reality, it is still too early to provide specific implementation details because no THz network has yet been built, let alone a multi-functional network. Nevertheless, we can foresee that joint algorithm and hardware design optimization for joint THz ultra-high-speed communications and advanced sensing, imaging, and localization paradigms will inevitably be a hot research topic in the near future. 



A typical THz communications system model is illustrated in Fig. \ref{Fig_RF}, where adaptive AoSAs at the transmitting side are configured to serve multiple receiving devices. Each subarray can be tuned to a specific frequency or can be assigned a specific modulation type. Furthermore, after DAC and before ADC, each subarray is fed by a dedicated RF chain. Note that very high-bandwidth low-resolution ADCs/DACs are required. Due to high directivity (``pencil'' beams), each subarray is effectively detached from its neighboring subarrays, and the role of baseband precoding is reduced to defining the utilization of subarrays, or simply to turning subarrays on and off. Depending on the desired communication distance, a required number of antenna elements per subarray is allocated for beamforming. Then, the number of possible subarray allocations, which is bounded by array dimensions and the number of RF chains, dictates the diversity gain. 




 

\subsection{Piggybacked Implementation}
\label{sec:Piggy}

For all the aforementioned reasons, sensing, imaging, and localization are expected to be executed at the same time over communication system nodes. It is thus intuitive to aim to have these applications piggybacked on THz communications by using the same space, time, and frequency resources, in the uplink, downlink, or both. For instance, in the case of electronic smelling, gaseous components can be identified by detecting variations in channel estimations regardless of the data being modulated over the carriers; only a few arbitrary carriers can result in good classification results (Fig. \ref{Fig_gas}). The same concept applies to imaging in reflection mode, in which reflection indices can be detected from received information-bearing carrier signals. Note that gas sensing and material detection can also be conducted simultaneously by assuming one, detecting the other, and then refining decisions over iterations. Nevertheless, a piggybacked implementation is much more convenient in a pulse-based system, where each modulated short-living pulse covers the entire THz band in the frequency domain, and can thus be used to conduct TDS-based sensing and imaging, as well as localization and navigation. 



\subsection{Dedicated Resources}
\label{sec:Dedicated}

Despite the attractiveness of full resource sharing, several factors make dedicated resource allocation schemes compulsory. For instance, the accuracy of carrier-based sensing and imaging increases by tuning carriers to absorption spectra that cannot carry information. While it is unlikely that frequencies much higher than $\unit[1]{THz}$ will be useful for communication purposes, sensing applications thrive at higher frequencies with congested absorption spectra. Hence, some applications naturally require separate frequency allocations. In fact, careful frequency planning is required to prevent the overlap of harmonic products when supporting multiple applications. Furthermore, this merging of applications often requires cooperation from nearby infrastructure or users, which is not always available. To distribute THz communication resources, the baseband precoder can assign each chain in Fig. \ref{Fig_RF} to a specific application. Alternatively, the same resources can be shared, but at dedicated time slots. 

\subsection{Role of Machine Learning}
\label{sec:ML}

Artificial intelligence is expected to play an important role in future THz systems, especially now that these systems are capable of generating a huge amount of data in a very rapid manner. From a communications perspective, machine learning can be used to enable higher spectrum utilization in UM-MIMO settings by replacing conventional channel estimation and channel coding schemes. Machine learning can also be used for beamforming and efficient data detection in generalized index modulation schemes. The importance of artificial intelligence is further emphasized in THz sensing and imaging, where feature extraction and classification can be conducted via principal component analyses and a support vector machine, or approximate entropy and a deep neural network, respectively. Furthermore, machine learning can prove to be of great importance in THz-based localization, for map interpolation and extrapolation as well as for cooperative localization and multi-source data fusion.

\subsection{Health and Privacy Concerns}
\label{sec:Concerns}

THz-band applications raise many health and privacy concerns. The International Commission on Non-Ionizing Radiation Protection (ICNIRP) considers heating to be the main risk factor of THz radiation. Since THz radiation does not penetrate the body, this risk is confined to the heating of skin tissues. Nevertheless, we can not tell for sure whether THz radiation can cause skin cancer, for example. Given the inherent densification and high antenna gains at THz, it is very important to conduct studies to further understand the true impact of THz radiation on health (\!\cite{Rappaport8732419} and references therein). Furthermore, a privacy concern arises in high-resolution sensing, imaging, and localization. With high beamforming gains and precise beamsteering capabilities, high-quality imaging can be conducted from a distance, perhaps via see-through imaging, as in THz-based airport scanners. This can occur at the user's end (if a user's equipment is hacked) or over the network (assuming network-centric positioning). Combining this with machine learning techniques, the privacy concern is undeniable. Such concerns should be addressed for both the software and the hardware. 





\section{Conclusion}
\label{sec:Conc}

In this paper, we present a holistic and progressive vision of the future of THz communications. Alongside communications, THz technology can bring significant advances to the areas of imaging, sensing, and localization. We detailed the peculiarities of each of these applications at the THz band by reviewing relevant works and illustrating proof-of-concept simulations. We argued that merging THz-band applications will guide wireless communication research far beyond 5G use cases, and we showed that multiple THz applications can be seamlessly realized in real-time. 

\ifCLASSOPTIONcaptionsoff
  \newpage
\fi


\section*{Biographies}
\footnotesize

\textbf{Hadi Sarieddeen} (S'13-M'18) received his B.E. degree (summa cum laude) in computer and communications engineering from Notre Dame University-Louaize (NDU), Lebanon, in 2013, and his Ph.D. degree in electrical and computer engineering from the American University of Beirut (AUB), Beirut, Lebanon, in 2018. He is currently a postdoctoral research fellow at KAUST. His research interests are in the areas of communication theory and signal processing for wireless communications.

\textbf{Nasir Saeed} (S'14-M'16-SM'19) received his Bachelor of Telecommunication degree from the University of Engineering and Technology, Peshawar, Pakistan, in 2009 and his Master's degree in satellite navigation from Polito di Torino, Italy, in 2012. He received his Ph.D. degree in electronics and communication engineering from Hanyang University, Seoul, South Korea in 2015. He is currently a postdoctoral research fellow at KAUST. His current areas of interest include cognitive radio networks, underwater optical wireless communications, dimensionality reduction, and localization.

\textbf{Tareq Y. Al-Naffouri} (M'10-SM'18) Tareq Al-Naffouri received his B.S. degrees in mathematics and electrical engineering (with first honors) from KFUPM, Saudi Arabia, his M.S. degree in electrical engineering from Georgia Tech, Atlanta, in 1998, and his Ph.D. degree in electrical engineering from Stanford University, Stanford, CA, in 2004. He is currently a Professor at the Electrical Engineering Department, KAUST. His research interests lie in the areas of sparse, adaptive, and statistical signal processing and their applications, localization, and machine learning.


\textbf{Mohamed-Slim Alouini} (S'94-M'98-SM'03-F'09) was born in Tunis, Tunisia. He received his Ph.D. degree in Electrical Engineering from Caltech, CA, USA, in 1998. He served as a faculty member at the University of Minnesota, Minneapolis, MN, USA, then in Texas A\&M University at Qatar, Doha, Qatar, before joining KAUST as a Professor of Electrical Engineering in 2009. His current research interests include the modeling, design, and performance analysis of wireless communication systems.

\end{document}